# A Prior-Guided Joint Diffusion Model in Projection Domain for PET Tracer Conversion

Fang Chen, Weifeng Zhang, Xingyu Ai, BingXuan Li, An Li, Qiegen Liu, *Senior Member, IEEE*

***Abstract*—Positron emission tomography (PET) is widely used to assess metabolic activity, but its application is limited by the availability of radiotracers. $^{18}$F-labeled fluorodeoxyglucose ($^{18}$F-FDG) is the most commonly used tracer but shows limited effectiveness for certain tumors. In contrast, 6-$^{18}$F-fluoro-3,4-dihydroxy-L-phenylalanine ($^{18}$F-DOPA) offers higher specificity for neuroendocrine tumors and neurological disorders. However, the complexity of its synthesis process and constraints on transportation time have limited its clinical application. Among different forms of raw data acquired by the scanner, sinogram is a commonly used representation in PET imaging. Therefore, modeling in projection domain enables more direct utilization of the original information, potentially reducing the accumulation errors during the image reconstruction process. Inspired by these factors, this study proposes a prior-guided joint diffusion model (PJDM) for transforming $^{18}$F-FDG PET sinograms into $^{18}$F-DOPA PET sinograms. During inference, an initial synthetic $^{18}$F-DOPA PET sinogram is first generated using a higher-order hybrid sampler. This sinogram is then degraded and serves as an additional condition to guide the iterative refinement process. Experimental results demonstrated that PJDM effectively improved both sinogram quality and the final synthetic outcomes. The code is available at: https://github.com/yqx7150/PJDM.***

## I. INTRODUCTION

POSITRON emission tomography (PET) is a crucial molecular imaging technology that is widely applied in oncologic, neurologic, and cardiovascular fields [1-4]. By injecting radiotracers, PET provides information on tissue metabolism and functional activity, facilitating early disease diagnosis, staging assessment, and treatment monitoring. Currently, the most widely used PET radiotracer in clinical practice is $^{18}$F-labeled fluorodeoxyglucose ($^{18}$F-FDG), which reflects glucose metabolic activity and plays a critical role in the detection and staging of various common malignancies, including cervical cancer, colorectal cancer, esophageal cancer, lung cancer, lymphoma, and head and neck squamous cell carcinoma [5]. Studies have demonstrated that $^{18}$F-FDG PET is valuable in cancer staging, recurrence assessment, and treatment response monitoring [6]. However, $^{18}$F-FDG PET is not suitable for all disease types. For example, it has limited uptake and low imaging sensitivity in prostate cancer, hepatocellular carcinoma, renal cell carcinoma, low-grade lymphoma, low-grade sarcoma, and well-differentiated neuroendocrine tumors (NETs). Moreover, the renal excretion and bladder accumulation of $^{18}$F-FDG may affect the diagnostic accuracy of urinary system tumors [7]. In recent years, another PET radiotracer, 6-$^{18}$F-fluoro-3,4-dihydroxy-L-phenylalanine ($^{18}$F-DOPA), has shown promising potential in neuroendocrine tumor and neurological disease research. $^{18}$F-DOPA PET specifically reflects the uptake of amino acid precursors by neuroendocrine tumors and has demonstrated potential applications in Parkinson's disease, schizophrenia, and attention-deficit hyperactivity disorder (ADHD) [8]. Studies have indicated that $^{18}$F-DOPA PET provides reliable quantitative metrics for monitoring disease progression in Parkinson's disease, aiding in early diagnosis and disease evaluation [9].

Although $^{18}$F-DOPA PET has high specificity in the detection of neuroendocrine tumors and neurological diseases, there are deficiencies such as complex synthesis processes and transportation time limitations. First, the synthesis process of $^{18}$F-DOPA is complex and technically demanding. It requires professional radiochemical knowledge and strict quality control procedures to ensure its purity and stability. Any synthesis deviation may lead to a decrease in the signal strength of PET images or an increase in background noise. Second, compared with the widely used $^{18}$F-FDG, $^{18}$F-DOPA faces more time limitations in transportation and administration, which limits its clinical application scope [10].

Recent advancements in computer vision and deep learning have introduced new possibilities for PET imaging. In particular, cross-modal medical image synthesis has emerged as a promising approach to improving interoperability between different PET radiotracers. Traditional cross-modal image synthesis methods include atlas-based matching, sparse coding, and conventional machine learning models. However, these approaches exhibit limitations in computational efficiency, adaptability, and stability [11-15]. With the evolution of deep learning, convolutional neural networks (CNNs), generative adversarial networks (GANs), and diffusion models have been extensively applied in medical image synthesis. For instance, Morbée *et al*. [16] leveraged a U-Net-based model for MRI-to-CT synthesis, enhancing image translation accuracy. Armanious *et al*.

This work was supported by National Natural Science Foundation of China (621220033, 62201193). (F. Chen and W. Zhang are co-first authors) (Corresponding authors: A. Li and Q. Liu).

F. Chen, X. Ai, Q. Liu and A. Li are with School of Information Engineering, Nanchang University, Nanchang, China ({aixingyu.aiden, liuqiegen, lian}@ncu.edu.cn; chenfang@email.ncu.edu.cn)

W. Zhang is with School of Jiluan Academy, Nanchang University, Nanchang, China (zhangweifeng@email.ncu.edu.cn)

B. Li is with Institute of Artificial Intelligence, Hefei Comprehensive National Science Center, Hefei, China (libingxuan@iai.ustc.edu.cn).

[17] introduced the MedGAN framework, which integrates adversarial loss and structural preservation loss to achieve high-quality medical image translation. Li *et al*. [18] proposed the frequency-guided diffusion model, which incorporates frequency-domain filtering to preserve anatomical structures, demonstrating superior structural retention in zero-shot modality translation learning tasks.

The above research progress indicates that deep learning techniques hold considerable potential in the field of cross-modal medical image synthesis, effectively overcoming the limitations of traditional methods. However, most existing studies are based on the image domain, and there is little research on modality translation between different tracers. Based on this, this study explores the transformation between sinograms, which offers both unique advantages and notable challenges. On one hand, sinograms represent the direct outputs of PET scanners before image synthesis, thus working with this data allows more direct utilization, potentially reducing error accumulation introduced during the synthesis process. On the other hand, sinogram translation poses significant challenges. Compared to the image domain, where spatial structures and support regions are more consistent, the projection domain exhibits larger variations due to the way projection data is collected. This domain gap increases the difficulty of effective learning and translation between tracers.

Inspired by these factors, this study proposes a prior-guided joint diffusion model (PJDM) to synthesize $^{18}$F-DOPA PET sinogram from $^{18}$F-FDG PET sinogram, as illustrated in Fig. 1. The model effectively captures complex cross-modal relationships and improves the quality of the final synthesized images.

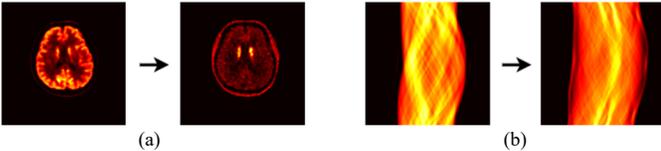

(a)                                         (b)

**Fig. 1.** PET tracer conversion between $^{18}$F-FDG and $^{18}$F-DOPA PET in image domain (a) and projection domain (b), respectively.

The proposed model employs a coarse-to-fine learning strategy, consisting of two stages. At the first stage, the model utilizes supervised learning with a limited number of paired samples to perform a coarse modality conversion, mapping $^{18}$F-FDG sinograms to $^{18}$F-DOPA sinograms. At the second stage, an unsupervised learning mechanism is introduced to leverage a large number of single-modality $^{18}$F-DOPA samples, thereby achieving fine-grained refinement.

The contributions in this study are summarized as follows:
- ***Training Framework Combining Supervised and Unsupervised Learning:*** In the initial stage, a limited set of paired $^{18}$F-FDG and $^{18}$F-DOPA data is used to train the coarse estimation model through supervised learning, enabling the fundamental transformation between tracer modalities. Subsequently, a larger-scale $^{18}$F-DOPA dataset is introduced, and an unsupervised learning strategy is employed to train the prior refinement model, which enhances the synthesize ability and improves the quality of the final $^{18}$F-DOPA PET image.
- ***Prior-Guided Joint Diffusion Model for Dual-Tracer Conversion in Projection Domain:*** The proposed model adopts a two-stage framework. In the first stage, a preliminary transformation from the $^{18}$F-FDG sinogram to the $^{18}$F-DOPA sinogram is achieved through a high-order hybrid sampler. In the second stage, the initially generated $^{18}$F-DOPA sinogram undergoes a degradation process involving Gaussian blurring, as well as random adjustments of contrast and brightness. This degraded sinogram is then used as additional prior information to guide the iterative refinement of the reverse process, ultimately generating high-fidelity $^{18}$F-DOPA PET sinograms.

The structure of this paper is organized as follows: Section II introduces the background knowledge of PET synthesis and diffusion models. Section III elaborates on the theoretical methodology of the proposed PJDM with detailed explanations. Section IV presents the experimental results, followed by further discussions and conclusions in Sections V and VI.

## II. PRELIMINARY

### A. PET Synthesis

PET provides metabolic information for comprehensive clinical decision-making. However, due to the limitations such as time, manpower, cost, and radiation exposure, acquiring complete multimodal imaging is not always feasible [19]. Consequently, medical image synthesis technology serves as a viable alternative. It aims to infer and generate target image modalities from existing information using computational methods, enabling cross-modal or within-modal image transformation [20].

Cross-modal transformation primarily involves predicting PET images from other modalities, such as MRI, to reduce dependency on PET scans. For instance, Hussein *et al*. [21] used an attention-based multi-scale 3D CNN to enhance cerebral blood flow measurement and diagnosis via multimodal MRI fusion, achieving high-quality MRI-to-PET conversion and cerebrovascular disease classification. Xie *et al*. [22] proposed a novel joint diffusion attention model that synthesizes PET images from high-field and ultra-high-field MRI using joint probability distributions and attention strategies. In contrast, within-modal transformation focuses on synthesizing and converting different image features within the same imaging modality to optimize the diagnostic value of PET images and improve clinical usability. For example, Xue *et al*. [23] proposed a GAN-based LCPR-Net that directly synthesizes full-count PET images from low-count sinogram data, improving image reconstruction quality and efficiency through cyclic consistency constraints and domain transformation operations. Hong *et al*. [24] proposed a spatial-temporal guided diffusion transformer probabilistic model, predicting delayed PET images from its initial scan using a U-Net and transformer framework.

This study focuses on a specific domain of within-modal transformation: the synthesis of one PET tracer image into another. In particular, in the fields of neuroimaging and oncology, $^{18}$F-DOPA and $^{18}$F-FDG are two radiotracers with distinct clinical applications. $^{18}$F-FDG is widely used for diagnosis of various cancers, although it is known to exhibit suboptimal imaging performance for neuroendocrine tumors. In contrast, $^{18}$F-DOPA is employed for the metabolic evaluation of the nervous system, especially in the diagnosis of neurodegenerative diseases such as Parkinson's disease. As a result, the conversion of $^{18}$F-FDG images into $^{18}$F-DOPA images is expected to provide more comprehensive information for lesion analysis.

### B. Diffusion Models

Diffusion models are generative models that have attracted much attention in the field of deep learning in recent years. The core mechanism is to gradually add noise to the data, making the data gradually disordered, and then restore the original data distribution by learning the reverse denoising process. Because it does not rely on labeled data, diffusion models have been widely applied in tasks such as image generation, data augmentation and denoising.

As shown in Fig. 2, diffusion models can be divided into supervised and unsupervised diffusion models according to whether a supervised signal is introduced. Supervised learning and unsupervised learning are two typical mechanisms in machine learning. Supervised learning directly learns the mapping between input and output using labeled data. Its main goal is to minimize prediction errors and enable the model to produce accurate outputs. In contrast, unsupervised learning is a data-driven approach that does not rely on labeled data but discovers the intrinsic structure of the data by learning the hidden representations within it.

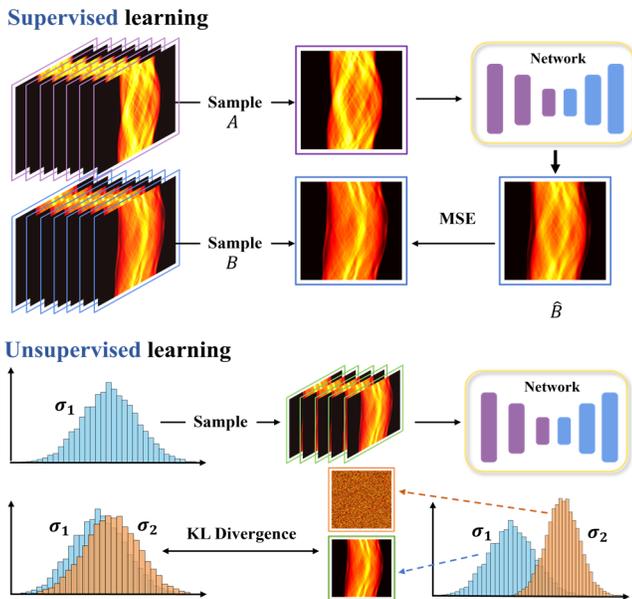

**Fig. 2.** The fundamental training paradigms of supervised and unsupervised learning.

Unsupervised diffusion models, such as denoising diffusion probabilistic models (DDPMs) [25] and score-based generative models [26], mainly focus on learning data distributions to generate high-quality samples. These models do not rely on labels but learn by leveraging the inherent characteristics of the data. However, in order to make the generated results meet the requirements of specific tasks, diffusion models can also integrate supervised learning by introducing conditional information [27].

A notable approach is Brownian bridge diffusion models (BBDM) [28], which achieves modality conversion between images by performing diffusion in the latent space. This method directly learns the correlations between the two modalities, and thereby improves image quality.

In this study, a supervised diffusion bridge model is integrated with a conditional diffusion probabilistic model based on the DDPM framework to enable high-quality transformations between PET images acquired with different tracers. The diffusion bridge model is designed to establish a bridge between two specific data distributions, enabling a deep neural network to directly learn transformations between the two modalities.

Meanwhile, the conditional diffusion probabilistic model gradually adds noise to the image through a forward process and employs a trained neural network to iteratively denoise the data. Building upon this, the model incorporates external conditional information to guide the generative process, guaranteeing that the generation process is subject to conditional constraints.

## III. METHOD

### A. Motivation

In PET imaging, the injected tracer emits positrons (e$^+$) that subsequently annihilate with nearby electrons (e$^-$), producing pairs of gamma photons traveling in opposite directions. These photons are simultaneously detected by opposing de-tectors and recorded using coincidence counting, forming a matrix or sinogram.

Each row of the sinogram corresponds to projection data at a specific angle and axial position. Subsequently, image reconstruction algorithms are employed to process the sinogram in order to recover the distribution of the radioactive tracer, thereby indirectly mapping the functional processes underlying the distribution of the positron-emitting radionuclide [29]. Since sinograms inherently encode the geometric and physical properties of projection data, optimizing them can effectively enhance the overall data quality after back-projection into the image domain.

Despite the increasing interest in image synthesis tasks, most existing methods still rely solely on either supervised or unsupervised learning. However, each learning paradigm has its inherent limitations. As shown in Fig. 3, the synthesized $^{18}$F-DOPA images generated by a single unsupervised or supervised model demonstrate these constraints. Neural networks possess powerful fitting capabilities. Kurt *et al.* [30] shows that a neural network with hidden layers can approximate any nonlinear function if it has enough parameters. At the same time, the increase in the number of parameters also raises higher demands on the size of the dataset. Sun *et al.* [31] points out that the performance of a network is positively correlated with the amount of training data. Using supervised learning requires a large amount of paired data to achieve relatively accurate predictions. However, in medical imaging field, due to issues such as radiation exposure and patient privacy, the acquisition of training data is often subject to many constraints. In PET imaging, it is rare to obtain training data in which the same patient is injected with different types of tracers, making it difficult for supervised learning to achieve good modality conversion performance. As a complementary approach, unsupervised learning does not rely on manual labels. Instead, it extracts and models features by exploring the intrinsic structure, distributional properties, or low-dimensional representations of the data, thus improving sample efficiency and model generalization. However, unlike supervised learning, unsupervised methods cannot directly infer or generate well-defined target images. Their primary function is to reveal the underlying structure of data, lacking explicit supervision for specific tasks. As a result, their effectiveness is often limited when applied to clearly defined tasks such as prediction or modality translation.

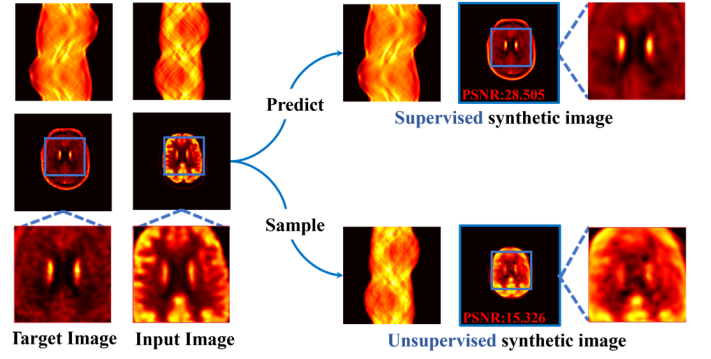

Fig. 3. The $^{18}$F-DOPA images synthesized using a single unsupervised or supervised diffusion model.

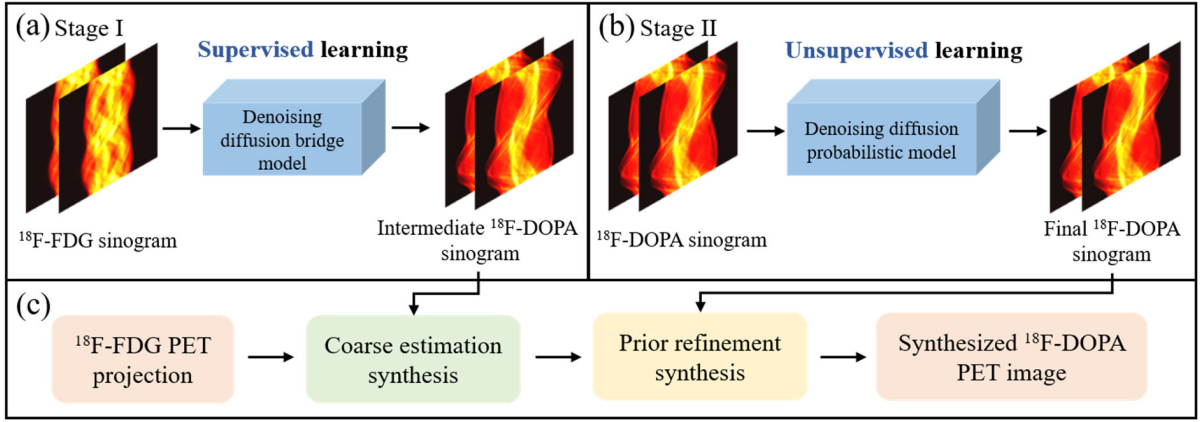

Fig. 4. Main process of PJDM. (a) Simple training process of the coarse estimation model, (b) Simple training process of the prior refinement model, (c) Simple testing process of PJDM.

Accordingly, this study proposes a cascaded mechanism that integrates supervised and unsupervised learning, as illustrated in Fig. 4. This mechanism consists of two stages. In the first stage, the model employs supervised learning with a limited number of paired samples to perform an initial modality conversion from $^{18}$F-FDG to $^{18}$F-DOPA. The second stage utilizes unsupervised learning with a large number of $^{18}$F-DOPA samples to further enhance the details and quality of the generated images. This approach allows the model to fully utilize the advantages of unsupervised learning while minimizing the need for supervised training samples, resulting in more detailed and accurate conversion outcomes.

### B. Coarse Estimation

The basic framework of PJDM is shown in Fig. 5. The first stage introduces a coarse estimation model to achieve an initial transformation from $^{18}$F-FDG sinograms to $^{18}$F-DOPA sinograms.

A diffusion process is constructed and represented by a set of time-indexed variables $\{x_t\}_{t=0}^{T}$, where the initial state satisfies $x_0 \sim q_{data}(^{18}\text{F-DOPA})$, and the terminal state satisfies $x_T \sim q_{data}(^{18}\text{F-FDG})$. This diffusion process is conditioned via Doob's $h$-transform to ensure that its endpoint belongs to the known $^{18}$F-FDG sinogram data distribution:

$$dx_t = [f(x_t,t) + g^2(t)h(x_t,t,x_T,T)]dt + g(t)z, \quad (1)$$
$$h(x_t,t,x_T,T) = \nabla_{x_t} \log p(x_T \mid x_t),$$

where $f: \mathbb{R}^d \times [0,T] \to \mathbb{R}^d$ is a vector-valued drift function, $g:[0,T] \to \mathbb{R}$ is a scalar-valued diffusion coefficient, and $z \sim \mathcal{N}(0,I)$ denotes the noise. The function $h(x_t,t,x_T,T)$ represents the log-gradient of the transition kernel generated by the original SDE from time $t$ to $T$, guiding the diffusion process toward the endpoint $y$.

The SDE in the reverse process is given by:

$$d(x_t,t)^{SDE} = [f(x_t,t) - g^2(t)(s(x_t,t,x_T,T) - h(x_t,t,x_T,T))]dt \quad (2)$$
$$+ g(t)z.$$

For the ODE process, a tunable parameter $w$ is introduced to control the strength of the drift term:

$$d(x_t,t)^{ODE} = [f(x_t,t) - g^2(t)((s(x_t,t,x_T,T)/2) - wh(x_t,t,x_T,T))]dt, \quad (3)$$
$$s(x_t,t,x_T,T) = \nabla_{x_t} \log q(x_t \mid x_T).$$

The time-reversed trajectory can be modified by adjusting $w$, enabling the generation of more diverse samples. In the training mechanism, supervised training is conducted between the paired $^{18}$F-FDG and $^{18}$F-DOPA datasets to achieve the initial modality conversion from $^{18}$F-DOPA sinograms to $^{18}$F-FDG sinograms.

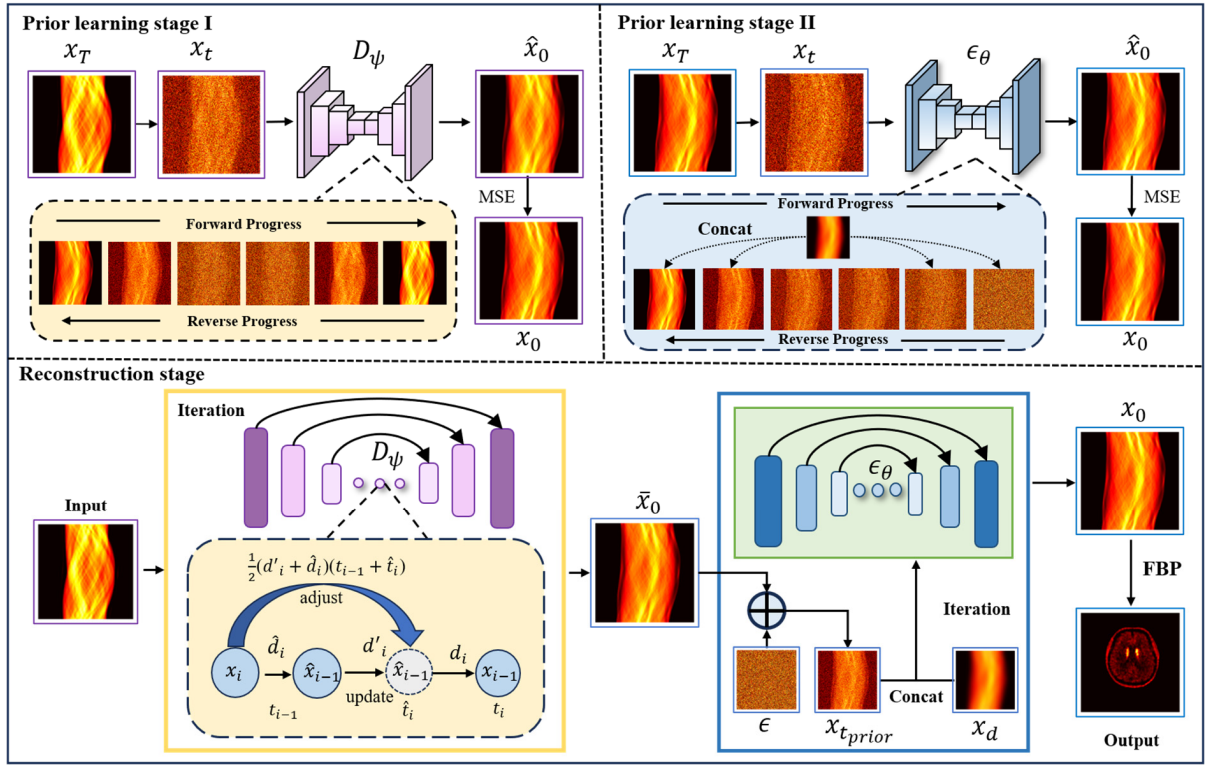

Fig. 5. The overall framework of PJDM. First, the coarse estimation model $D_\psi$ synthesizes an initial $^{18}$F-DOPA sinogram from the $^{18}$F-FDG sinogram while preserving the spatial structure. Then, the $^{18}$F-DOPA sinogram, after undergoing Gaussian blurring along with random contrast and brightness variations, is used as additional information to guide the prior refinement model $\epsilon_\theta$, facilitating detail restoration and enhancing the synthesis quality.

By designing $f(x_t,t)$ and $g^2(t)$, transition kernel of the underlying diffusion process can be obtained, leading to the explicit expression for $h(x_t,t,x_T,T)$. The sampling process which is shown in Eq. (2) and Eq. (3) requires approximation of the score $\nabla x_t \log q(x_t | x_T)$.

However, the true score function cannot be explicitly solved. Therefore, a neural network is applied to approximate the true score function by matching a tractable quantity using the denoising score matching approach.

Furthermore, the output of neural network $F_\psi$ can be parameterized by $D_\psi(x_t,t) = c_{skip}(t)x_t + c_{out}(t)F_\psi(c_{in}(t)x_t, c_{noise}(t))$ to further calculate the score function $\nabla x_t \log q(x_t | x_T)$ of target distribution [32].

Thus, by giving data $(x_0, x_T) \sim q_{data}(^{18}F\text{-DOPA}, ^{18}F\text{-FDG})$, $x_t \sim q(x_t | x_0, x_T)$, $t \sim p(t)$, where $p(t)$ is a nonzero time sampling distribution defined over $[0,T]$, our optimization objective is to minimize the following loss function:

$$\mathcal{L}(\psi) = \min \mathbb{E}_{x_t,x_T,t}[w(t) \| D_\psi(x_t,t) - x_0 \|^2], \quad (4)$$

in this case, $x_t = a_t x_T + b_t x_0 + \sqrt{c_t}\epsilon$ for $\epsilon \sim \mathcal{N}(0,I)$. Let $a_t = \alpha_t / \alpha_T * SNR_T / SNR_t$, $b_t = \alpha_t(1 - SNR_T / SNR_t)$, as well as $c_t = \sigma_t^2(1 - SNR_T / SNR_t)$, where $\alpha_t$ and $\sigma_t$ represent the predefined schedules for the signal and noise. The signal-to-noise ratio SNR at time $t$ is given by $SNR_t = \alpha_t^2 / \sigma_t^2$. Additionally, $w(t)$ is an arbitrary nonzero function to calculate the loss weight.

During the iterative synthesis stage, the proposed PJDM model first achieves the initial modality conversion. Based on the higher-order hybrid sampler proposed by Karras et. al [32], the sampling steps are discretized as $t_N > t_{N-1} > \cdots > t_0$, with each time step having progressively decreasing intervals. Additionally, a scaling hyperparameter $m$ is introduced, which defines the step size ratio between $t_i$ and $t_{i-1}$. The time interval $\hat{t}_i = t_i + m(t_{i-1} - t_i)$ is used for the Euler-Maruyama step, as described below:

$$\begin{aligned} d_i &= -d(x_i,t_i)^{SDE} \\ \hat{x}_i &= x_i + d_i(\hat{t}_i - t_i). \end{aligned} \quad (5)$$

By parameterizing the model output that predicts $x_0$ as $D_\theta(x_t,t)$, the bridge score $\nabla x_t \log q(x_t | x_T) = s(x_t,t,x_T,T)$ can be further predicted. As for the interval $[t_{i-1}, t_i - m(t_i - t_{i-1})]$, it is used for the Heun step:

$$\begin{aligned} \hat{d}_i &= -d(\hat{x}_i,\hat{t}_i)^{ODE} \\ x_{i-1} &= \hat{x}_i + \hat{d}_i(t_{i-1} - \hat{t}_i). \end{aligned} \quad (6)$$

Then, using the second-order correction [25], the explicit trapezoidal method is applied at $t_{i-1}$ to obtain the next $x_{i-1}$:

$$\begin{aligned} d'_i &= -d(x_{i-1}, t_{i-1})^{ODE} \\ x_{i-1} &= \hat{x}_i + (d'_i/2 + \hat{d}_i/2)(t_{i-1} - \hat{t}_i). \end{aligned} \quad (7)$$

The forward SDE process transforms the complex data distribution containing $^{18}$F-DOPA information into the known $^{18}$F-FDG distribution by injecting noise. The noise-perturbed data is then processed along the reverse trajectory. By estimating the score function, the reverse process combines SDE and ODE to generate samples. Thus, after iterative sampling, the modality conversion from $^{18}$F-FDG is performed.

### C. Prior Refinement

After performing a preliminary modality translation from $^{18}$F-FDG to $^{18}$F-DOPA sinogram using the coarse estimation model, a conditional diffusion model is further introduced to enhance the representation of structural details.

In the forward process, the noise is progressively injected into the original image and the formulation is given as follows:

$$q(x_t \mid x_0) = \mathcal{N}(x_t; \sqrt{\bar{\alpha}_t} x_0, (1-\bar{\alpha}_t)I), \quad (8)$$

where $\alpha_t$ represents the noise scheduling parameters, and $\beta_t = 1 - \alpha_t$, where $\beta_t \in (0,1)$ increases monotonically with the timestep $t \in \{1,\ldots T\}$. Let $\bar{\alpha}_t = \prod_{i=1}^{t} \alpha_i$, as $t$ increases, $\bar{\alpha}_t$ decreases and approaches zero when $t$ becomes sufficiently large.

Since the task is aimed at refining the initially generated $^{18}$F-DOPA sinograms to enhance fine details, an effective strategy is adopted, in which a blurred, contrast and brightness altered DOPA sinogram $x_d$ can be represented by:

$$x_d = R(x_0), \quad (9)$$

where $R$ is defined as the degradation function. Through this conditional information, the generation process is guided accordingly. Then $x_t$ and $x_d$ are used as a pair of inputs to the model. The corresponding optimization objective is defined as:

$$\min \mathbb{E}_{t,x_0,\epsilon}[\|\epsilon - \epsilon_\theta(concat(\sqrt{\bar{\alpha}_t} x_0 + \sqrt{1-\bar{\alpha}_t}\epsilon, x_d), t)\|^2], \quad (10)$$

where $\epsilon_\theta$ denotes the neural network-based prediction of the noise $\epsilon \in \mathcal{N}(0,I)$ introduced during the forward process.

During the reverse process, the preliminary DOPA sinogram generated in the first stage is subjected to blurring, random contrast, brightness adjustments, and subsequently utilized as conditional information. To reduce the amount of noise introduced during generation, the reverse process starts at an intermediate timestep $t_{prior}$. This approach preserves the modeling capacity of the diffusion process while effectively retaining structural information, thereby facilitating the generation of higher quality DOPA sinograms. The inference process can be expressed as:

$$p(x_{t_{prior}-1} \mid x_t, c) = \mathcal{N}(x_{t_{prior}-1}; \mu_\theta(x_{t_{prior}}, c), \sigma_{t_{prior}}I), \quad (11)$$

$$\mu_\theta(x_{t_{prior}}, c) = \frac{1}{\sqrt{\alpha_{t_{prior}}}}(x_{t_{prior}} - \frac{1-\alpha_{t_{prior}}}{\sqrt{1-\bar{\alpha}_{t_{prior}}}}\epsilon_\theta(x_{t_{prior}}, c)), \quad (12)$$

where $c$ denotes the spatial condition embedded into the model, and the formulation $c = concat(x_t, x_d)$ represents the concatenation of paired images along the channel dimension using the concatenate operation. Thus, at each synthesis step, the noisy image $x_t$ is concatenated with the conditional image $x_d$ along the channel dimension.

Therefore, the prediction of $x_{t-1}$ from $x_t$ can be finally formulated as:

$$x_{t_{prior}-1} = \frac{1}{\sqrt{\alpha_{t_{prior}}}}(x_{t_{prior}} - \frac{1-\alpha_{t_{prior}}}{\sqrt{1-\bar{\alpha}_{t_{prior}}}}\epsilon_\theta(c, t_{prior})) + \sigma_{t_{prior}} z, \quad (13)$$

where $z \sim \mathcal{N}(0,I)$ is the noise.

Algorithm 1 outlines the two stage PJDM process: a supervised diffusion model first performs coarse FDG-to-DOPA conversion, followed by an unsupervised refinement stage that uses degraded priors to enhance structural details via conditional diffusion.

---

**Algorithm 1: Prior-Guided Joint Diffusion Model (PJDM)**

**Synthesis stage I**

**Require**: Time steps $\{t_i\}_{i=0}^{N}$, maximum time step $T$, step ratio $m$ and guidance strength $w$.

**For** $i = N, \cdots, 1, 0$ **do**
    $x_N \sim q_{data}(^{18}\text{F-FDG})$, $\epsilon_i \sim \mathcal{N}(0,I)$
    $\hat{t}_i = t_i + m(t_{i-1} - t_i)$
    $d_i = -d(x_i, t_i, x_N, T)^{SDE}$, according to Eq. (2)
    $\hat{x}_i = x_i + d_i(\hat{t}_i - t_i)$
    $\hat{d}_i = -d(\hat{x}_i, \hat{t}_i, x_N, T)^{ODE}$, according to Eq. (3)
    $x_{i-1} = \hat{x}_i + \hat{d}_i(t_{i-1} - \hat{t}_i)$
    **if** $i \neq 1$ **then**
        $d_i' = -d(x_{i-1}, t_{i-1}, x_N, T)^{ODE}$, according to Eq. (3)
        $x_{i-1} = \hat{x}_i + (d_i'/2 + \hat{d}_i/2)(t_{i-1} - \hat{t}_i)$
    **end if**
**end for**
**Return** $\hat{x}_0$

**Synthesis stage II**

**Require**: Intermediate timestep $t_{prior}$, coarse prior $\hat{x}_0$ from synthesis stage I.

$x_d = R(\hat{x}_0)$
$x_t \sim q(x_t \mid \hat{x}_0)$, $t = t_{prior}$
**For** $t = t_{prior}, \cdots, 1, 0$ **do**
    $z \in \mathcal{N}(0,I)$ if $t > 1$, else $z = 0$
    $c = concat(x_t, x_d)$
    $x_{t-1} = \frac{1}{\sqrt{\alpha_t}}(x_t - \frac{1-\alpha_t}{\sqrt{1-\bar{\alpha}_t}}\epsilon_\theta(c,t)) + \sigma_t z$
**end for**
**Return** $x_0$

---

## IV. EXPERIMENTS

### A. Data Specification

In this study, the dataset was collected from the Department of Neurology at the First Affiliated Hospital of Sun Yat-sen

University. Between 2020 and 2023, the hospital recruited a sequence of patients diagnosed with idiopathic Parkinson's disease (PD) based on the criteria established by the British Parkinson's Disease Association Brain Bank. Patients with a history of dementia, stroke, abnormal brain structure, encephalitis, a poor response to levodopa (<30%), disease duration of less than five years, or an inability to adhere to the study protocol were excluded from the study.

The dataset comprised PET scans from 196 patients, among which 2,643 paired $^{18}$F-FDG and $^{18}$F-DOPA PET scans were selected to train the coarse estimation model, enabling it to learn the distributional mapping between $^{18}$F-FDG and $^{18}$F-DOPA sinograms. In addition, 3,877 unpaired $^{18}$F-DOPA PET scans were used to train the prior refinement model, further enhancing the model's ability to capture fine structural details in the $^{18}$F-DOPA sinograms. Furthermore, an independent test set comprising 141 paired scans from an all-digital PET system (Brain PET B320 platform by RAYSOLUTION Healthcare Co., Ltd.) was employed to validate and assess the final performance of the model. It is important to note that this study was conducted with the approval of the local ethics committee and with informed consent from all patients.

### B. Experimental Setup

The prior-guided joint diffusion model proposed in this study was implemented using the PyTorch framework and trained on an NVIDIA GeForce RTX 4080 GPU with 16 GB of memory. The AdamW optimizer was employed, with an initial learning rate of $1 \times 10^{-4}$ for the coarse estimation model and $5 \times 10^{-5}$ for the prior refinement model. The coarse estimation model was developed based on the denoising diffusion bridge model (DDBM) framework [33], while the prior refinement model was constructed upon the DDPM architecture. During the synthesis of $^{18}$F-DOPA sinograms, the maximum number of sampling steps for the coarse estimation model was set to $T=40$. In the prior refinement stage, a fast sampling strategy based on DDIM [34] was adopted. At this stage, the initially generated $^{18}$F-DOPA sinograms were first degraded using Gaussian blurring along with random contrast and brightness perturbations, to simulate image degradation. The resulting degraded image served as prior information to guide the subsequent iterative refinement process, with prior guidance introduced at step $t_{prior}=185$. This strategy effectively enhanced the image quality and structural fidelity of the generated results.

### C. Quantitative Indices

To evaluate the quality of the synthesized data, this study employs three metrics for quantitative assessment: peak signal-to-noise ratio (PSNR), structural similarity index (SSIM), and normalized root mean square error (NRMSE).

PSNR describes the relationship between the maximum possible power of a signal and the power of noise corruption. Higher PSNR indicates better synthesis quality. Let $I$ and $I'$ represent the reconstructed and ground truth images, respectively. The PSNR is defined as:

$$PSNR(I, I') = 20\lg(\max(I)/\|I - I'\|_2), \quad (14)$$

where $\max(I)$ is the maximum pixel value of the image.

SSIM measures the similarity between the reconstructed and ground truth images, and is calculated as:

$$SSIM(I, I') = \frac{(2\mu_I \mu_{I'} + c_1)(2\sigma_{II'} + c_2)}{(\mu_I^2 + \mu_{I'}^2 + c_1)(\sigma_I^2 + \sigma_{I'}^2 + c_2)}, \quad (15)$$

where $\mu_I$ and $\sigma_I^2$ are the mean and variance of image $I$, $\sigma_{II'}$ is the covariance between $I$ and $I'$. $c_1$ and $c_2$ are constants to stabilize the division.

NRMSE normalizes the error to improve comparability across different datasets or scales. It is calculated as:

$$NRMSE = (\sqrt{\sum_{i=1}^{n}(I'_i - I_i)^2/n})/(\max(I') - \min(I')), \quad (16)$$

where $n$ is the total number of pixels.

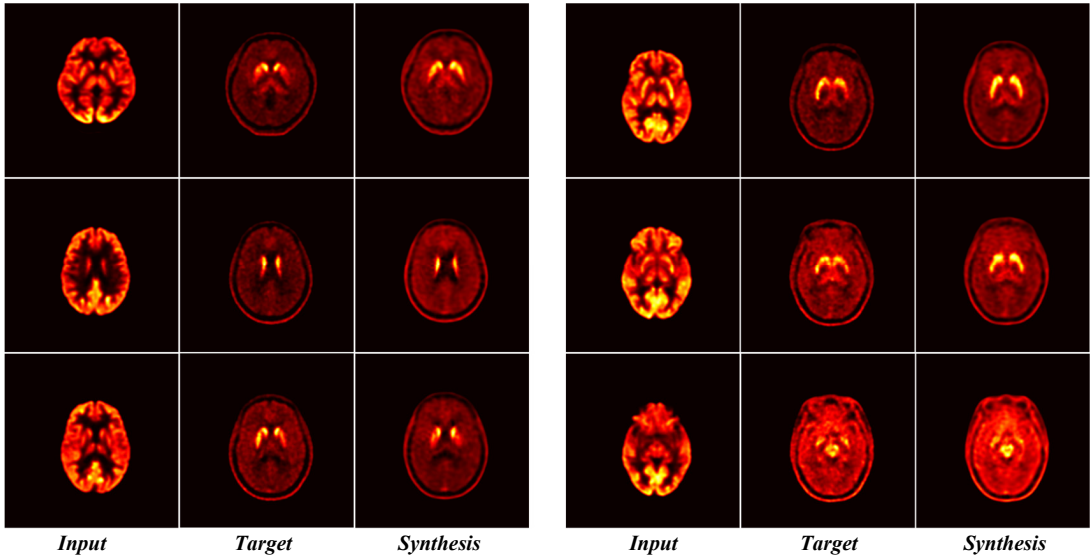

**Fig. 6.** The synthesized results of PJDM. The first column on the left shows the input image ($^{18}$F-FDG), the second column presents the target image ($^{18}$F-DOPA), and the third column displays the synthesized $^{18}$F-DOPA image produced by the model.

### D. Synthesis Experiments

*Imaging Result Comparisons:* To validate the effectiveness of the proposed algorithm, qualitative and quantitative analyses were conducted to assess the quality of the generated images. Fig. 6 presents the experimental results, where the first column shows the input images, the second column displays the target $^{18}$F-DOPA PET images, the third column presents the generated $^{18}$F-DOPA PET images. Overall, the generated images exhibit high similarity to the reference images in terms of overall contours and internal structural information. This indicates that the proposed method can accurately approximate the overall morphology and structural characteristics of $^{18}$F-DOPA images.

Several established approaches are used to evaluate the performance of PJDM, including U-Net [35], CycleGAN [36], Pix2Pix [37] and cold diffusion (CD) [38]. Table I summarizes the average quantitative evaluation metrics on the test dataset, including peak signal-to-noise ratio (PSNR), structural similarity index (SSIM), and normalized root mean square error (NRMSE). The results demonstrate that PJDM achieves the best performance across all three metrics, with a PSNR of 24.98 dB, SSIM of 0.812, and an NRMSE of 0.064. Notably, PJDM outperforms the second-best method, CD, by 1.12 dB in PSNR and 0.059 in SSIM, highlighting its capability in generating high-quality $^{18}$F-DOPA images.

TABLE I
COMPARISON OF TRACER CONVERSION UNDER DIFFERENT METHODS

| Method | PSNR↑ | SSIM↑ | NRMSE↓ |
|---|---|---|---|
| U-Net [35] | 23.23 | 0.684 | 0.128 |
| CycleGAN [36] | 22.12 | 0.704 | 0.082 |
| Pix2Pix [37] | 22.69 | 0.647 | 0.085 |
| CD [38] | 23.86 | 0.753 | 0.068 |
| **PJDM** | **24.98** | **0.812** | **0.064** |

Meanwhile, the qualitative analysis presented in Fig. 7 further supports that the images generated by PJDM maintain a high structural consistency with the target $^{18}$F-DOPA images. While some edge contours appear slightly smoother compared to the reference images, the overall shape is well preserved. In contrast, images generated by U-Net suffer from insufficient sharpness and excessive smoothing, which may impact lesion interpretation. Additionally, CycleGAN and Pix2Pix methods retain certain $^{18}$F-FDG-specific structural features in the generated $^{18}$F-DOPA images, resulting in incomplete structural transformation and compromising image quality. As for CD, although the overall contours of the generated DOPA images are consistent with the reference images, there are significant differences in internal structures, and the generated details lack accuracy.

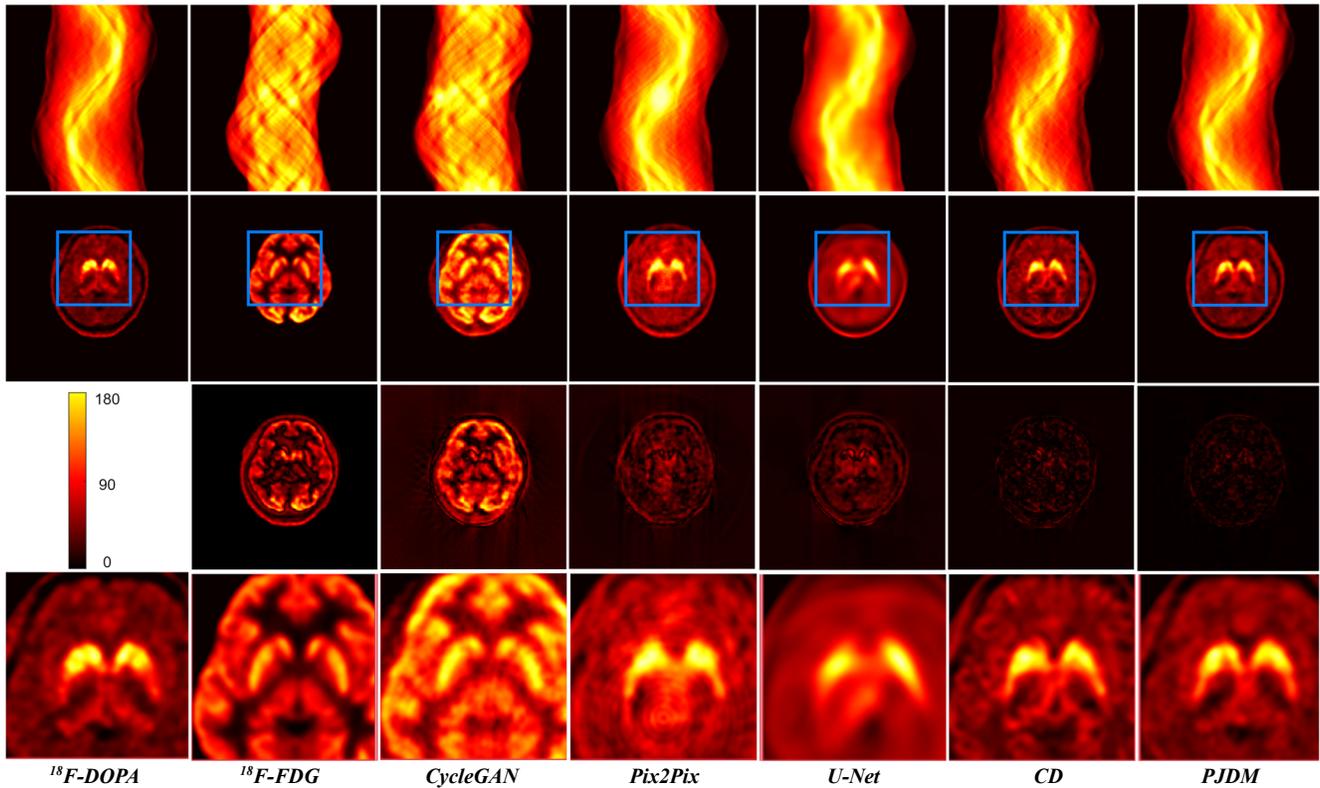

Fig. 7. Comparison of synthesized results using different methods. The first row shows the sinograms, the second row presents the synthesized images, the third row displays the residuals, and the last row presents the zoomed-in images of key regions.

The profile lines of striatum region are illustrated in Fig. 8. Notably, PJDM exhibits superior visual consistency in this area, with its profile line most closely matching that of the reference $^{18}$F-DOPA images, while other methods present more apparent discrepancies. This highlights the effectiveness of PJDM in preserving critical structures and accurately capturing the characteristics of $^{18}$F-DOPA images, thereby enabling more precise image transformation.

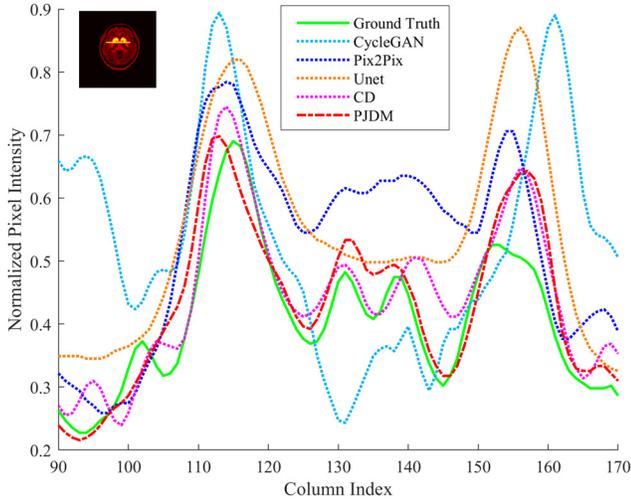

**Fig. 8.** The profile lines of the striatum region. The results indicate that the recovery values obtained by PJDM are closest to the ground truth.

In addition, a dedicated analysis is conducted on the synthesized sinogram results. In Fig. 9, a comparison is presented to illustrate the performance of different methods in sinogram synthesis. The figure is organized into two rows: the first row shows sinograms generated by various methods, while the second row displays magnified regions to facilitate clearer observation of fine details. In the image domain, distinct characteristics are exhibited by different methods during the sinogram synthesis process. Noticeable blurring in the edge regions is observed in the sinogram synthesized by the U-Net method. In the cases of CycleGAN and Pix2Pix, certain structural features of the original $^{18}$F-FDG sinogram are retained during the transformation process, resulting in incomplete structural conversion. Although overall contour consistency with the target $^{18}$F-DOPA sinogram is achieved by the CD method, significant differences remain in the internal structure, and the synthesized details are found to lack accuracy. In contrast, the sinogram synthesized by the proposed method most closely matches the target $^{18}$F-DOPA in terms of overall contour and demonstrates high consistency in internal structural details. These results suggest that the proposed method is the most effective in preserving key structures and accurately capturing the characteristic features of the $^{18}$F-DOPA sinogram.

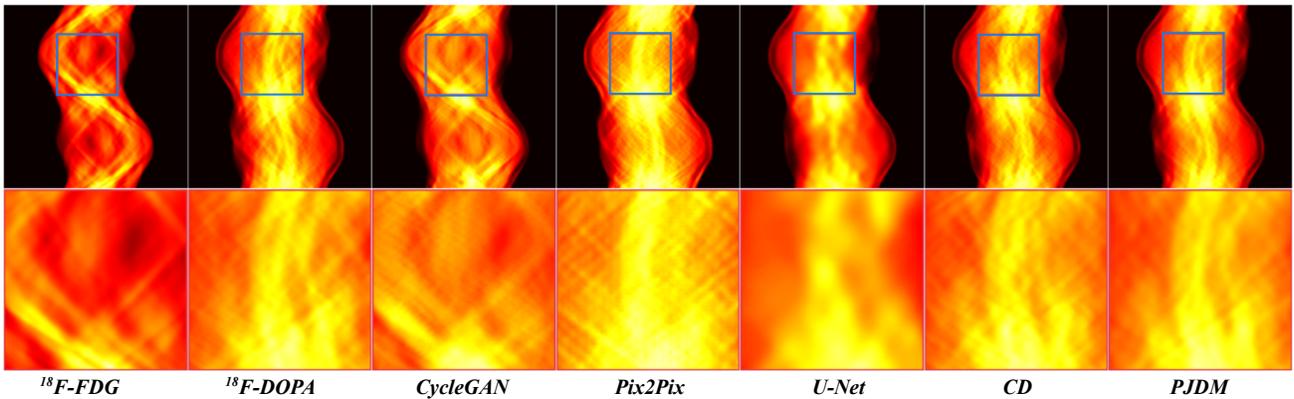

$^{18}$F-FDG  $\quad$  $^{18}$F-DOPA  $\quad$  CycleGAN  $\quad$  Pix2Pix  $\quad$  U-Net  $\quad$  CD  $\quad$  PJDM

**Fig. 9.** Comparison of sinogram synthesis results using different methods. The first row shows the synthesized sinograms, while the second row presents the corresponding zoomed-in regions.

*Ablation Studies:* To evaluate the effectiveness of each component, ablation studies are conducted. The individual impact of each module on performance is independently analyzed to ensure a comprehensive and accurate assessment. As shown in Table II, the ablation results demonstrate that the combination of the coarse estimation (CE) model and the prior refinement (PR) model significantly improves the quality of image synthesis. Quantitative metrics indicate that the CE+PR configuration achieves notable gains compared to using either model alone: PSNR increases from 24.25 to 24.98, SSIM improves from 0.751 to 0.812, and NRMSE decreases from 0.081 to 0.064. These results suggest that incorporating the prior refinement module effectively enhances synthesized image quality.

The above conclusion is further supported by the qualitative analysis in Fig. 10. When the CE model is used alone, the synthesized images exhibit less detailed structures and show noticeable noise in the surrounding areas. When the PR model is used alone, it can achieve a basic translation from $^{18}$F-FDG to $^{18}$F-DOPA image. However, there are significant differences in detail compared to the ground truth.

TABLE II
QUANTITATIVE COMPARISON OF TRACER CONVERSION METRICS AMONG DIFFERENT MODULES

| Metric | (w/o) CE | (w/o) PR | PJDM |
|---|---|---|---|
| PSNR↑ | 24.25 | 22.87 | **24.98** |
| SSIM↑ | 0.751 | 0.739 | **0.812** |
| NRMSE↓ | 0.081 | 0.091 | **0.064** |

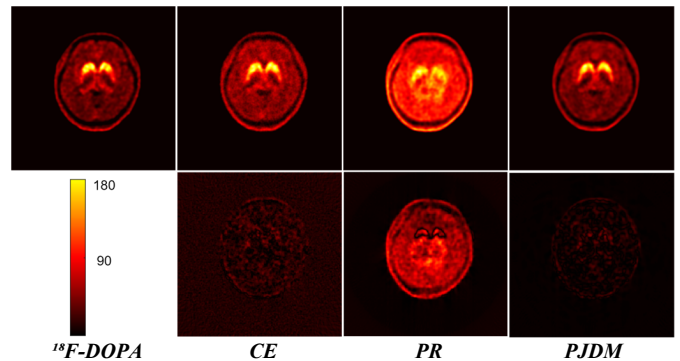

$^{18}$F-DOPA  $\quad$  CE  $\quad$  PR  $\quad$  PJDM

**Fig. 10.** The comparison results of ablation studies. The details of the image enhanced after applying the PR method.

The synthesized images obtained with the CE+PR combination demonstrate the highest structural fidelity and detail completeness, with the smallest residual errors and the best visual quality. In summary, compared to using a single model alone, the integration of CE and PR effectively enhances the quality of synthesis. Additionally, the edge preservation performance of each module is evaluated by comparing the degree of alignment between the generated contours and the ground truth contours.

As shown in Fig. 11, the profile line of PJDM method achieves the highest accuracy and the closest alignment with the ground truth. In contrast, solely PR model exhibits the poorest consistency, while CE model ranks second in performance. Notably, the PJDM method demonstrates significantly better contour consistency in the striatum region, further confirming its outstanding and stable capability in preserving crucial details.

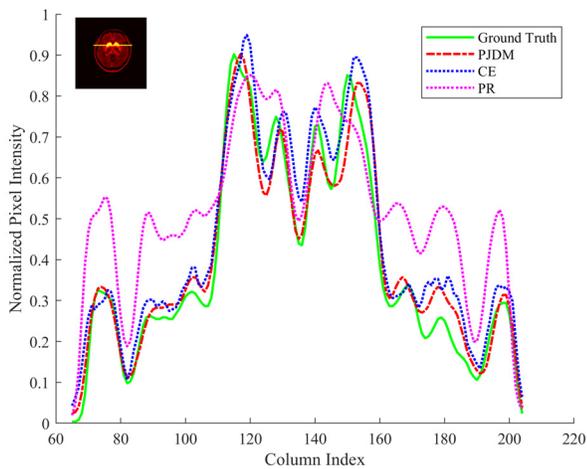

**Fig. 11**. The profile comparison after removing different modules. It shows that the results obtained using both CE and PR are the closest to the ground truth.

## V. DISCUSSION

The proposed PJDM enhances the detail representation of $^{18}$F-DOPA PET images during the unsupervised learning phase by progressively removing Gaussian noise in the reverse process. Compared to traditional diffusion models that rely solely on Gaussian noise, alternative degradation strategies such as mosaic, Gaussian blurring, and pixelation could also be explored to further improve model performance.

Moreover, tuning hyperparameters such as the learning rate may further enhance the synthesis capability of PJDM. Future research can expand toward multimodal PET image synthesis, extending beyond the conversion between $^{18}$F-FDG and $^{18}$F-DOPA to include other tracers such as $^{18}$F-NaF and $^{18}$F-FLT. These advancements are expected to provide more comprehensive and accurate imaging solutions for various types of tumors, neurological disorders, and other medical research applications.

## VI. CONCLUSION

This study proposed a deep learning-based PJDM to achieve cross-modal transformation from $^{18}$F-FDG PET sinograms to $^{18}$F-DOPA PET sinograms. The cascaded model was trained using a coarse-to-fine strategy. In the first stage, supervised learning with paired data was employed to capture the fundamental mapping relationship between $^{18}$F-FDG and $^{18}$F-DOPA. In the second stage, unsupervised learning was introduced, where the initially synthesized $^{18}$F-DOPA sinograms were subjected to Gaussian blurring along with random contrast and brightness perturbations to generate prior information. This prior guided the reverse process iteratively to further refine the quality of the synthesized images. Experimental results demonstrated that the proposed method not only preserved structural information but also enhanced details. Future work will focus on optimizing the generative capability of the diffusion model and exploring its potential in multi-tracer transformation.